\documentclass[pdflatex,sn-mathphys-num]{sn-jnl}

\usepackage{tikz}
\usetikzlibrary{patterns,patterns.meta} 
\usepackage{bbold}
\usepackage{tikzscale}
\usepackage{hyperref}

\usepackage{graphicx}%
\usepackage{multirow}%
\usepackage{amsmath,amssymb,amsfonts}%
\usepackage{amsthm}%
\usepackage{mathrsfs}%
\usepackage[title]{appendix}%
\usepackage{xcolor}%
\usepackage{textcomp}%
\usepackage{manyfoot}%
\usepackage{booktabs}%
\usepackage{algorithm}%
\usepackage{algorithmicx}%
\usepackage{algpseudocode}%
\usepackage{listings}%

\theoremstyle{thmstyleone}%
\theoremstyle{thmstyletwo}%
\theoremstyle{thmstylethree}%
\begin{document}

\title[discrete probability]{A framework for general discrete probability calculations}

\author*[1,]{\fnm{Kacper} \sur{Topolnicki}}\email{kacper.topolnicki@uj.edu.pl}
\author[1]{\fnm{Roman} \sur{Skibiński}}\email{roman.skibinski@uj.edu.pl}

\affil[1]{\orgdiv{Institute of Physics}, \orgname{Jagiellonian University}, \orgaddress{\street{Prof. St. Łojasiewicza 11}, \city{Kraków}, \postcode{30-348}, \country{Poland}}}


\abstract{Probability follows a simple and concise set of rules.  
In practice, however,
reasoning about probability may be highly unintuitive and 
this leads to the possibility of miscalculations even for simple problems. 
We present an approach to facilitate the correct description of systems 
governed by discrete probability.
The methods described in this 
paper allow making general statements about the values of 
probability. This information can next
be used
to calculate any probability related to the system described by the statements. 
In case there are too few statements to provide a unique 
answer, entropy maximization is used to fill in the missing information.
The approach has wide potential applications that range from experimental physics to natural language analysis. 
The theoretical framework described here is also the basis of a software 
implementation 
that is designed to be straightforward to extend and 
requires only popular \emph{python} libraries, \emph{numpy},
\emph{torch}, \emph{sympy}, and \emph{scipy}.
}


\maketitle


\section{Introduction}


The goal of the approach described in this paper is to provide a simple tool 
for working with conditional probabilities for
discrete distributions.
We describe a theoretical framework that is used in the new \emph{ProbabilityIter} 
({\bf Probability} {\bf Iter}ator)
\emph{Python} library \cite{TopolnickiGitHub}.
Many types of probability calculations can be carried 
out using available software, for example \emph{Python's} \emph{pymc} \cite{pymc2023}
or \emph{pyro} \cite{pyro_joss}. These are large software packages
and focus mainly on continuous distributions. 
Many phenomena are, however, well described by discrete probabilities
and this is the focus of the described approach.
The framework is designed to be straightforward and the implementation
\cite{TopolnickiGitHub} 
can be easily embedded
in other software solutions. The \emph{Python} library \cite{TopolnickiGitHub} 
has few dependencies and uses only popular python tools: \emph{numpy}, \emph{pytorch}, 
\emph{sympy}, and \emph{scipy}.
The implementation is a direct reflection of the theoretical approach
described in this paper, making it possible to easily modify and extend
the code. The solutions are constructed from linear equations and the 
relevant matrices are exposed in the code for use in other algorithms
including machine learning applications.

An important feature of the framework
is flexibility. It accepts very general statements about discrete probability in 
the form:
\begin{itemize}
	\item[] $P(a_{1} | b_{1}) = \alpha_{1}$
	\item[] $P(a_{2} | b_{2}) = \alpha_{2}$
	\item[] $P(a_{3} | b_{3}) = \alpha_{3}$
	\item[] $\ldots$
\end{itemize}
where $a_{1,2,3,\ldots}$ and $b_{1,2,3,\ldots}$ are
logical statements and $\alpha_{1}, \alpha_{2}, \alpha_{3}, \ldots$
are values of the probability $P$.
After processing, this input allows the calculation of
probabilities $P(u | v)$ where $u, v$ are arbitrary probability statements
constructed from the same literals as $a_{1,2,3,\ldots}$ and 
$b_{1,2,3,\ldots}$. 
This approach has wide potential applications that range from experimental physics,
a more detailed example is given in Subsection \ref{ssol}, to natural language analysis,
for example the linter program \cite{SikoraGitHub}.
To the best of our knowledge there is a lack of software solutions
in this area and the library \cite{TopolnickiGitHub} is a new option.

The paper is organized as follows. Section \ref{logic} describes our choice 
to use a probability formalism that is based on logic. Section \ref{sdnf}
illustrates how general logical statements are processed. Section \ref{cons}
describes how each user input is translated to a single row
of a linear equation. Section \ref{max} discusses a typical scenario
where the number of user inputs are not sufficient to produce a unique 
result. Section \ref{aions} discusses potential applications and Section \ref{so} contains the summary.

\section{Probability based on logic}
\label{logic}

This paper uses a probability formalism based on logic, an
approach formalized in the Cox theorem
\cite{cox1961algebra} and later popularized in
\cite{jaynes2003probability}. In this picture, probability is always
conditional. The notation:
\begin{equation}
	P(a | b)
\label{prob1}
\end{equation}
will be interpreted as the probability of the logical proposition $a$ being true 
given that $b$ is true. In \cite{cox1961algebra} the rules
governing probability are derived from three assumptions or desiderata \cite{jaynes2003probability}:
probability is represented as a real number, probability calculations are in agreement with \textquotedblleft common sense\textquotedblright, 
and are consistent i.e. if some probability can be derived in more than one way then each method must
produce the same value. \textquotedblleft Common sense\textquotedblright\,in this context, means that probability values change 
continuously with new information becoming available,
see also \cite{Smith1989}. These three assumptions are made more precise in \cite{cox1961algebra,jaynes2003probability,Smith1989} and 
used to derive rules governing probability that are utilized in further parts of this paper.

Two basic properties related to the probability of disjunctions and the probability of 
conjunctions will be used for deriviations further in the text.
If $a$, $b$, and $c$ are logical propositions then:
\begin{equation}
P(a + b | c) = P(a | c) + P(b | c) - P(a b | c)
\label{sum_rule}
\end{equation}
where, following \cite{jaynes2003probability}, 
we use the notation $a + b \equiv a \lor b$ for the disjunction or sum
and $a b \equiv a \land b$ for the conjunction or product. The second rule 
is related to the conjunction:
\begin{equation}
P(a b | c) = P(a | b c) P(b | c) = P(b | a c) P(a | c).
\label{product_rule}
\end{equation}
Note that a set of corresponding
rules may be derived from the Kolmogorov axioms \cite{kolmogorov1950foundations}.
Calculations based on these corresponding rules will produce the same results for
events and their probabilities, but
we chose to work with logical propositions because we consider this language to be 
more natural in the context of Bayesian probability.

Finally, we introduce the logical proposition $x$ that reflects 
implicit prior knowledge about a system governed 
by probability. In principle $x$ could be a tautology, but
the introduction of this symbol is beneficial 
because it symbolizes the fact that it is very rare 
to approach a problem without initial assumptions.
For this reason in the following we will consistently multiply (conjugate) 
the right hand argument of $P(\ldots|(\ldots) x)$ with $x$.

\section{Disjunctive normal form}
\label{sdnf}

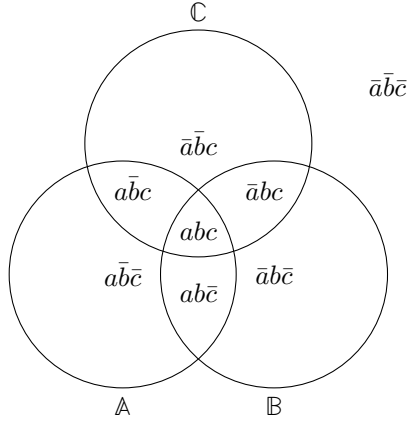
\begin{figure}[h!]
	\centering
	\begin{tikzpicture}
		\draw(-1,0) circle (1.5);
		\draw(1,0) circle (1.5);
		\draw(0,{sqrt(3)}) circle (1.5);
		\node[below] at (-1 , -1.5) {$\mathbb{A}$};
		\node[below] at (1 , -1.5) {$\mathbb{B}$};
		\node[above] at (0 , {sqrt(3) + 1.5}) {$\mathbb{C}$};
		\node at (0 , {sqrt(3) / 3.0}) {$a b c$};
		\node at (-1 , 0) {$a \bar{b} \bar{c}$};
		\node at (1 , 0) {$\bar{a} b \bar{c}$};
		\node at (0 , {sqrt(3)}) {$\bar{a} \bar{b} c$};
		\node[below] at (0 , 0) {$a b \bar{c}$};
		\node[above left] at (-0.5 , {0.5 * sqrt(3)}) {$a \bar{b} c$};
		\node[above right] at (0.5 , {0.5 * sqrt(3)}) {$\bar{a} b c$};
		\node at (2.5 , 2.5) {$\bar{a} \bar{b} \bar{c}$};
	\end{tikzpicture}
	\caption{Illustration of the DNF for the 
	statements: $a \equiv z \in \mathbb{A}$, $b \equiv z \in \mathbb{B}$, 
	$c \equiv z \in \mathbb{C}$ where $z$ is a point on the plane. The placement of the
    $z$ point on this plane
	divides the space into $8$ non intersecting regions. In each region only one of the $8$
    logical expressions, marked in the illustration, is true.}
	\label{example}
\end{figure}

Calculations related to probability can be simplified if propositions
are written in Disjunctive Normal Form (DNF). For illustrative purpouses,
consider a point $z$ on a plane and 
three logical propositions $a$, $b$, $c$ as illustrated in Figure \ref{example}.
The three propositions determine if the point belongs to set $\mathbb{A}$,
$\mathbb{B}$, $\mathbb{C}$ respectively.
Looking at this diagram, it is apparent that any statement regarding the value of $z$ may be written 
as a disjunction of $8$ statements from the set:
\begin{equation}
\text{DNF}(\{a , b , c\}) = \{
		a b c,
		\bar{a} b c,
		a \bar{b} c,
		a b \bar{c},
		\bar{a} \bar{b} c,
		a \bar{b} \bar{c},
		\bar{a} b \bar{c},
		\bar{a} \bar{b} \bar{c}
\}
\label{dnf}
\end{equation}
where the bar is used for negation. For example, $b$ may be written as
$a b c + \bar{a} b c + a b \bar{c} + \bar{a} b \bar{c}$ and $a c$ as
$a \bar{b} c + a b c$. 

This observation can be generalized. Any logical statement $v$ constructed of literals $\mathbb{S} = \{a, b, c, \ldots\}$
can be written as
\begin{equation}
	v = \sum_{d \in \text{DNF}(\mathbb{S})} v_{d} d
\label{dnf_1}
\end{equation}
where $v_{d}$ is $1$ (true) if $d$ is in the DNF expansion of $v$ and $0$ (false) otherwise.
More precisely we will be using the so called full DNF
where $\text{DNF}(\mathbb{S})$ contains all possible conjunctions and each literal, 
positive or negative,
appears exactly once in every conjunction. Using this property, we may write:
\begin{equation}
	P(v | x) = P\left(\sum_{d \in \text{DNF}(\mathbb{S})} v_{d} d \Bigg| x\right) = \sum_{d \in \text{DNF}(\mathbb{S})} v_{d} P(d | x).
	\label{sum_rule_1}
\end{equation}
where the meaning of $v_{d}$ depends on the context. The value of $v_{d}$ inside $P$
is either true or false and the value $v_{d}$ outside $P$ is the real number equivalent: $1$ or $0$ respecively.
The last equality in \eqref{sum_rule_1} holds because the conjunction of any two different expressions in $\text{DNF}(\mathbb{S})$ 
is false and the probability of a false statement being true must be $0$. 
As a consequence the last term on the right hand side of \eqref{sum_rule} 
is canceled out and only a single sum remains.
The sum rule \eqref{sum_rule} and product rule \eqref{product_rule} can now be generalized and 
simplified using \eqref{sum_rule_1}. If we introduce another logical proposition in 
DNF form:
\begin{equation}
	u = \sum_{d \in \text{DNF}(\mathbb{S})} u_{d} d
\label{dnf_2}
\end{equation}
then:
\begin{equation}
	P(u + v | x) = P(\sum_{d \in \text{DNF}(\mathbb{S})} (u_{d} + v_{d}) d | x) = 
	\sum_{d \in \text{DNF}(\mathbb{S})} \min(1 , u_{d} + v_{d}) P(d | x),
	\label{sum_rule_2}
\end{equation}
and
\begin{equation}
	P(u v | x) = P(\sum_{d \in \text{DNF}(\mathbb{S})} u_{d} v_{d} d | x) = 
	\sum_{d \in \text{DNF}(\mathbb{S})} u_{d} v_{d} P(d | x).
	\label{product_rule_1}
\end{equation}
Again, the values $u_{d}$, $v_{d}$ are considered to have two different interpretations.
Inside $P$ they are treated as logical values and outside of $P$ they 
are interpreted real numbers. The factor $\min(1 , u_{d} v_{d})$ 
in the last sum in \eqref{sum_rule_2}
has the value $1$ 
only if the logical value $u_{d} v_{d}$ is true and $0$ otherwise.
This means that the last sum in \eqref{sum_rule_2} is effectively 
over those conjunctions that appear in the DNF form
of $u$ or the DNF form of $v$
while the last sum in \eqref{product_rule_1} is over those conjunctions that
appear both in the DNF expansion of $u$ and the DNF expansion of $v$.
These properties can be intuitively understood by considering 
the probabilities of disjointed sets as in Figure \ref{example}
and suggest a representation of the probability as a vector with
each coordinate associated with the probability of a single conjunction in $\text{DNF}{(\mathbb{S})}$. 

Using the example in Figure \ref{example}, any probability associated with this system can be calculated 
from the vector:
\begin{equation}
	p = \begin{pmatrix}
		P(a b c | x) \\
		P(\bar{a} b c | x) \\
		P(a \bar{b} c | x) \\
		P(a b \bar{c} | x) \\
		P(\bar{a} \bar{b} c | x) \\
		P(a \bar{b} \bar{c} | x) \\
		P(\bar{a} b \bar{c} | x) \\
		P(\bar{a} \bar{b} \bar{c} | x) 
	\end{pmatrix}.
	\label{prob}
\end{equation}
The goal of our system is to calculate or estimate this vector and use it to 
work out arbitrary probabilities requested by the user.

\section{Constructing the linear system}
\label{cons}

Our framework expects the user to supply the logical expressions $u$, $v$, 
and the probability value $\alpha$.
This input will be interpreted as statements about probability in the form:
\begin{equation}
	P(u | v x) = \alpha.
	\label{input}
\end{equation}
Multiple input statements of this form taken together will be used to work out the 
vector of probabilities. 
Each input is translated into a single row of the
augmented matrix $(A|b)$ of a linear system
\[
	A p = b
\]
whose solution $p$ is a vector of probabilities, for example \eqref{prob}. Note
that some user inputs may result in a linear system that does not have solutions.
The problematic input can, however, be found by checking the system matrix after 
adding each row and the implementatinon of these checks is left to the users of \cite{TopolnickiGitHub}.

Using the product rule \eqref{product_rule}
the input \eqref{input}
may be equivalently be written as:
\begin{equation}
	\frac{P(u v | x)}{P(v | x)} = \alpha.
\label{input1}
\end{equation}
In the special case where $v$ is true the denominator in \eqref{input1} can be dropped
as a consequence of normalization. The numerator can also be simplified resulting in the following equation:
\begin{equation}
	P(u | x) = \alpha 
\end{equation}
which can be rewritten using \eqref{dnf_2} as:
\begin{equation}
	\sum_{d \in \text{DNF}(\mathbb{S})} u_{d} P(d | x) = \alpha.
	\label{eq1}
\end{equation}
A single row of $A$ can be directly transcribed from the left hand of this equation
and the corresponding element of $b$ can be directly read from the right hand side.

In the more general case where the $v$ is not explicitly true (has the truth value $1$) equation \eqref{input1}
can be rewritten in the following way:
\begin{equation}
	P(u v | x) - \alpha P(v | x) = 0.
\end{equation}
Using \eqref{product_rule_1} this can be expanded as:
\begin{equation}
	\sum_{d \in \text{DNF}(\mathbb{S})} (u_{d} v_{d} - \alpha v_{d}) P(d | x) = 0
	\label{eq2}
\end{equation}
Again, the left hand side of this equation can be transcribed to a single row of $A$,
the corresponding element in $b$ is $0$.

Using equations \eqref{eq1} and \eqref{eq2} each user input is transcribed into a row of $(A|b)$.
In practice, working with these equations is made easier by utilizing the \emph{SymPy}
\emph{Python} library \cite{10.7717/peerj-cs.103} and our code makes extensive use of the capabilities
of this library. 
With the solution of the linear equation $A p = b$ we can work out the values of
probabilities $P(u | v x)$ where $u$, $v$ are arbitrary
logical statements in the system using the property: 
\begin{equation}
P(u | v x) 
= \frac{P(u v | x)}{P(v | x)} 
= \frac{\sum_{d \in \text{DNF}(\mathbb{S})} u_{d} v_{d} P(d | x)}
{\sum_{d \in \text{DNF}(\mathbb{S})} v_{d} P(d | x)},
\label{numdem}
\end{equation}
where the values of $P(d|x)$ can be read from the vector $p$.
Please note additionally that the number of equations resulting from user input will typically
be insufficient to find a unique solution to $A p = b$ and we will address this problem in Section \ref{max}.

\subsection{Computational complexity}

Expanding $\mathbb{S}$ by a new literal results in the doubling of the number of elements in $\text{DNF}(\mathbb{S})$.
Since the number of elements is $|\text{DNF}(\mathbb{S})| = 2^{|\mathbb{S}|}$,
the size of vectors introduced in Section \ref{cons}, and the number of columns in $A$ grows 
quickly with the number of literals in $\mathbb{S}$. For example, to represent a system with only $|\mathbb{S}| = 34$
literals with a vector of double precision numbers would require over $100$ GB of memory
and storing a matrix that acts on this vector would be difficult on most modern hardware.
As a result the practical use of our code would
be limited to systems describable by a small number of logical propositions.

To mitigate this problem,
the user may enter a number of logical statements that play the role of constraints.
It is then assumed that the conjugation of these user constraints, $c$,
must be true.
When expanded into (full) DNF form $c = \sum_{d \in \text{DNF}(\mathbb{S})} c_{d} d$ and 
the set $\text{DNF}_{c}(\mathbb{S}) \equiv \{d : d \in \text{DNF}(\mathbb{S}) , c_{d} = 1\}$ may 
be used to replace $\text{DNF}(\mathbb{S})$ in all calculations. 
Depending on the quantity and scope of the constraints the number of elements 
in $\text{DNF}_{c}(\mathbb{S})$ may be significantly smaller then $\text{DNF}(\mathbb{S})$
leading to reduced computational complexity. In practice, the elements in $\text{DNF}_{c}(\mathbb{S})$
are calculated using the \emph{Sympy} library \cite{10.7717/peerj-cs.103}.


\section{Obtaining the solution and maximizing entropy}
\label{max}

We assume that in a typical use case the number of user entered equations is not enough to make 
the solution to the linear system introduced in Section \ref{cons} unique.
In order to have more control over the available solutions, we will be considering an alternative
equation:
\begin{equation}
	A p = \alpha b
	\label{sol2}
\end{equation}
where we introduce an additional real parameter $\alpha$. If $\alpha = 1$ then \eqref{sol2} is 
equivalent to $A p = b$.
Equation \eqref{sol2} may be rewritten in the form of a null space equation:
\begin{equation}
	(A | -b) 
	\begin{pmatrix}
		p \\
		\alpha 
	\end{pmatrix}
	\equiv M y = 0 
	\label{sol1}
\end{equation}
where the dimension of the vector $y$ is one greater then the dimension of $p$. 
The additional parameter $\alpha$ is superfluous but allows the reformulation 
of the problem as \eqref{sol1} and
solutions to 
this equation are unique up to a constant factor which may always be chosen so that the last component
$\alpha = 1$.

The choice to use the form \eqref{sol1} was partially motivated by the fact that many
\emph{Python} numerical libraries, including libraries designed for machine learning,
implement the calculation of null vectors. However, this form also has additional advantages. If 
\begin{equation}
	y^{(1)}, y^{(2)}, \ldots , y^{(N)}
\label{solts}
\end{equation}
are solutions to
\eqref{sol1} then any linear combination of these vectors can be made into a solution to the original
$A p = b$. The linear combination only needs to be multiplied by a constant factor so that the last
component is $1$. In principle, this means that an optimization procedure could search the space of linear
combinations for a solution with the highest entropy. In practice, this could
fail because the sign
of the components of the linear combination of vectors \eqref{solts} is unconstrained.

If the components of the vector are intended to be interpreted as probability when $\alpha = 1 > 0$ then 
a linear combination of \eqref{solts} is a candidate for a solution
only if all components of have the same sign, positive or negative.  
These two possibilities are connected by multiplication by a factor $-1$ which 
allows us to focus only linear combinations with all positive
components.

The first step in obtaining the correct solution of equation \eqref{sol1} involves finding a single
linear combination of \eqref{solts} whose components are all positive. 
This combination can be calculated using linear optimization 
with the following setup \cite{5128410}:
\begin{itemize}
	\item Find $y^{(+)} \in \mathbb{R}^{D + 1}$ where $D$ is the number of elements 
		in $\text{DNF}_{c}(\mathbb{S})$ such that:
	\begin{enumerate}
		\item The vector $- \sum_{j} y^{(+)}_{j}$ is maximized.
		\item The conditions $y^{(+)}_{j} > 1$ for all $j$, and $\tilde{M} y^{(+)} = 0$ are satistied.
	\end{enumerate}
\end{itemize}
The rows of $\tilde{M}$ contain vectors that form the basis of the orthogonal complement of the span of solutions \eqref{solts}.
The linear optimization problem is solved using the \emph{scipy} library \cite{2020SciPy-NMeth} 
\emph{scipy.optimize.linprog} function. 

Once $y^{(+)}$ is known it is possible to construct a search space
to finding the entropy maximizing solution and we chose the following approach.
The solution is parametrized by $\gamma^{(1)} , \gamma^{(2)}, \ldots, \gamma^{(N)}$ real numbers that correspond
to the $N$ vectors in \eqref{solts} and by an additional real number parameter $\beta$ that corresponds to the calculated
positive solution $y^{(+)}$. Given $\beta^{2} y^{(+)}$, a vector with all positive components, a range for 
$\delta^{(1)}$, $\delta^{(1)}_{\text{min}} \le \delta^{(1)} \le \delta^{(1)}_{\text{max}}$, is calculated so that the components of
$\beta^{2} u + \delta^{(1)} y^{(1)}$ all remain positive. 
A single value from this range
is chosen using the formula 
\begin{equation}
	\delta^{(1)} = \delta^{(1)}_{\text{min}} + \frac{1}{2}(\sin(\gamma^{(1)}) + 1.0) (\delta^{(1)}_{\text{max}} - \delta^{(1)}_{\text{min}})
	\label{range}	
\end{equation}
where the unconstrained value $\gamma^{(1)}$ is transformed into $\delta^{(1)}$ in the range 
$\delta^{(1)}_{\text{min}} \le \delta^{(1)} \le \delta^{(1)}_{\text{max}}$.
Next given $\beta^{2} y^{(+)} + \delta^{(1)} y^{(1)}$
a range for $\delta^{(2)}$ is calculated so that the components of 
$\beta^{2} y^{(+)} + \delta^{(1)} y^{(1)} + \delta^{(2)} y^{(2)}$ are all positive and a value $\delta^{(2)}$ calculated
from $\gamma^{(2)}$ using a transoformation analogous to \eqref{range}. This procedure
is repeated until the allowable ranges for all $\delta^{(1)} , \delta^{(2)}, \ldots, \delta^{(N)}$
are known and the vector $\beta^{2} y^{(+)} + \delta^{(1)} y^{(1)} + \delta^{(2)} y^{(2)} + \ldots \delta^{(N)} y^{(N)}$
has all positive components. Finally this vector can be normalized in such a way that the last coordinate is $1$
and the vector of probabilities, $p$, read off from the previous coordinates. The entropy 
\[
	S = \sum_{i} -p_{i} \ln{p_{i}}
\]
can be calculated from these probabilities and
backpropagation can be used to calculate the gradient of the entropy with respect to the unconstrained parameters 
$\gamma^{(1)} , \gamma^{(2)}, \ldots, \gamma^{(N)}, \beta$. 
This information may be appropriately used
in consecutive iterations of the optimization procedure,
for example using the \emph{Adam} optimizer \cite{DBLP:journals/corr/KingmaB14},
to modify the parameters 
$\gamma^{(1)} , \gamma^{(2)}, \ldots, \gamma^{(N)}, \beta$
and arrive at an entropy maximizing solution.

\begin{figure}[H]
	\centering
	\includegraphics[width = 0.75 \textwidth]{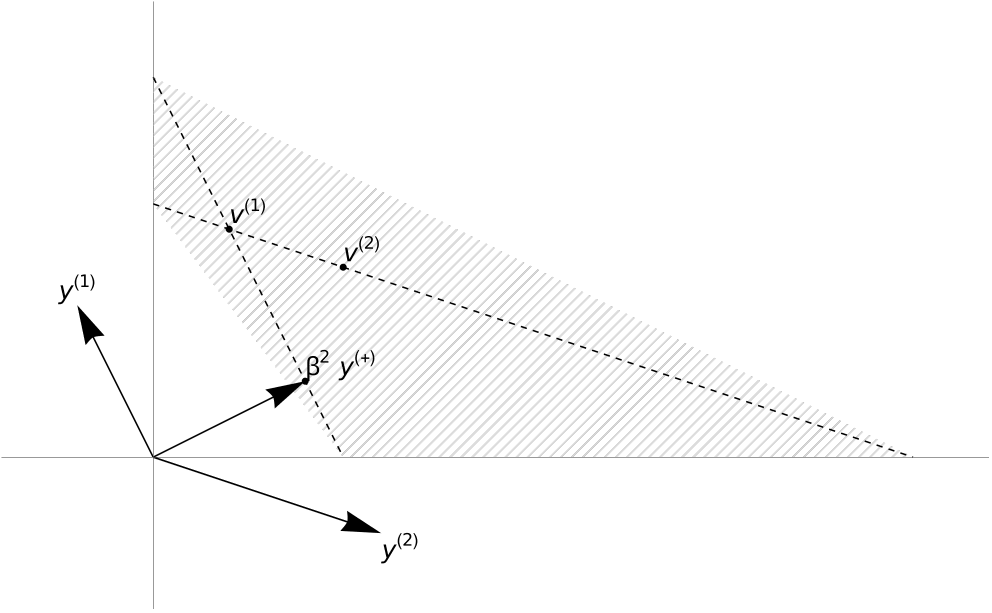}
	\caption{In a single iteration in two dimensions, 
	probing solutions to find an entropy maximizing 
	vector of probabilities starts with a solution to \eqref{sol1} that has
	all positive coordinates, for example $\beta^{2} y^{(+)} = y^{(1)} + y^{(2)}$. 
	Next the space is expanded to the hatched region: First a range of $\gamma^{(1)}$
	is found so that the vector $v^{(1)} = \beta^{2} y^{(+)} + \gamma^{(1)} y^{(1)}$ has all positive components;
	finally a range of $\gamma^{(2)}$ is calculated so that the vector
	$v^{(2)} = \beta^{2} y^{(+)} + \gamma^{(1)} y^{(1)} + \gamma^{(2)} y^{(2)}$ has all positive components.
	The real parameter $\beta$ may be modified to expand the search space to the entire first quadrant.
	}
	\label{ill}
\end{figure}

There are many ways to chose positive solutions. The approach described here
has the benefit of giving the parameters of the optimization a systematic meaning.
It is illustrated in Figure \ref{ill} for the two dimensional case.


\section{Applications}
\label{aions}

The approach described in the previous sections has wide potential applications that range 
from natural language analysis to experimental
physics. An interesting application, 
related to natural language
analysis, is the linter program \cite{SikoraGitHub}. 
Logical, in the
Aristotelian sense, errors are relatively easy to spot in a text
by experienced journalists. Erroneous statements, in the context of probability,
are much more difficult to find.
The linter \cite{SikoraGitHub} allows the user to reveal statements in the text that
may be misleading in this sense.
It is written 
using \cite{TopolnickiGitHub} and allows
the user to select fragments of natural language text that
constitute logical statements. Annotations can be used to attach 
probability values to these statements, and the linter program can use 
these annotations to answer questions related to probability.

In the following, we discuss a hypothetical problem related to experimental physics.
This scenario presents an opportunity to make unjustified inferences.
Avoiding these types of errors in real scenarios, especially since real experiments are more complicated than the 
hypothesised situation, may lead to significant cost savings.
Subsection \ref{ssol} shows how the approach described in the previous sections
is used in \cite{TopolnickiGitHub} to obtain a unique probability value that
is used in deciding a crucial experimental step.


\subsection{Obtaining a unique solution}
\label{ssol}

To demonstrate the framework and test the library \cite{TopolnickiGitHub}, let us propose the following hypothetical experimental
scenario. Three samples, labeled $1$, $2$, and $3$ are to be tested for the existence of a new 
unstable nuclear element, and only one of the three samples contains this element.
There exists an apparatus capable, with high certainty, of determining whether
a sample contains this new element, however, operating this instrument is
expensive and the experimental budget must be carefully planned.

The testing apparatus has three sample inlets. Only the 
sample in the first inlet
will be tested for the existence of the new element. The instrument 
has, however, the additional capability of revealing a sample from the other two inlets
that does not contain the new element after running a pre-test.
The experimental team's budget allows them to run the pre-test once and the
full test on the sample in port one once.
Having no additional information, the experimental team initially planned to run the full test
on the sample labeled $1$. 
Pre-testing revealed that the sample labeled $2$ does not to contain the new element.
The experimental team must 
decide whether this new
determination should influence their choice of sample on which to run the main test.

This scenario is, of course, equivalent to the well known Monty Hall problem
originally solved by Marilyn vos Savant, see for example \cite{mh}.
A careful consideration of the probabilities
describing the hypothetical experimental scenario results in the probability $\frac{1}{3}$ 
of the new element being in the sample labeled $1$ and the probability $\frac{2}{3}$ 
of the new element being in the sample labeled $3$. Taking this into account, the rational
choice for the experimental team is to change their initial choice and run the main test on the sample
labeled $3$. The naive approach is to assume that no additional information is gained and 
to test the sample labeled $1$ as originally planned. Depending on the cost of 
running the test on this sample, ignoring the information about the
sample labeled $2$ might lead to a potentially 
costly mistake. 

It might seem that physicists, due to their methodological training, would be
immune to making the wrong, naive, decision in this hypothetical scenario.
However, the Monty Hall problem started a debate in which many
academics, including members of the scientific community and mathematicians, were arguing for an incorrect
answer that ignores the additional information \cite{granberg2014monty}. 
The scenario described above may seem abstract, but similar decisions are frequently made by experimental physicists
making the library \cite{TopolnickiGitHub} a valuable tool.
Below we demonstrate how the probabilities
in the hypothetical experimental scenario can be calculated using \cite{TopolnickiGitHub}.

After following the instructions in \cite{TopolnickiGitHub},
the code below can be used to calculate the correct probability values using the
{\bf piter} \emph{Python} module. We will describe each line of this code and 
show how it relates to the problem: 
\begin{lstlisting}[language=Python , numbers = left]
from piter import Piter
from sympy.abc import A , B , C , O
from sympy import true
import numpy as np

p = Piter({A , B , C , O})
p.addConstraint((A & ~B & ~C) | (~A & B & ~C) | (~A & ~B & C))
p.addConstraint(~(O & B))
p.addP(A , true , 1.0 / 3.0)
p.addP(B , true , 1.0 / 3.0)
p.addP(C , true , 1.0 / 3.0)
p.addP(O , A , 0.5)
p.addP(O , C , 1.0)
p.finalize()

ab = p.getNumpy()

a = ab[: , :-1]
b = ab[: , -1]

x , residuals , rank , s = np.linalg.lstsq(a , b)

num , dem = p.getNumDem(A , O)
dont_switch = np.sum(num * x) / np.sum(dem * x)

print("prob. of success when testing sample 1 : " , dont_switch)

num , dem = p.getNumDem(C , O)
switch = np.sum(num * x) / np.sum(dem * x)

print("prob. of success when switching to sample 3:" , switch)
\end{lstlisting}
and results in the following output:
\begin{lstlisting}
prob. of success when testing sample 1 :  0.33333333333333326
prob. of success when switching to sample 3: 0.6666666666666667
\end{lstlisting}

Lines $1-4$ contain import statements. In line $1$ the \emph{Piter}
class is imported. In the following line four symbols are
imported from the \emph{sympy} library. Here, $A$, $B$, and $C$ symbolize statements that the new element is in sample $1$, $2$ and $3$ respectively. The symbol $O$ stands for the statement
that, after pre-testing, the sample labeled $2$ was revealed 
to not contain the new element. In line $3$ the sympy \emph{true}
value is imported, and it will be used on the right 
side of $P(\ldots|\ldots)$ in conditional probability to represent
a situation without initial assumptions (except $x$). Finally 
line $4$ imports the \emph{numpy} library.

Lines $6-14$ contain the main logic, we will use the 
the notation and problem setup from \cite{mh}. In line $6$ an
instance of \emph{Piter} is created to represent a universe
for probability calculations where statements can be constructed
from the four literals $A$, $B$, $C$, and $O$. Next in lines
$7-8$ constraints are added. The first constraint
means that only one of three possible situations is possible: the new
element is in the sample labeled $1$ (\verb|A & ~B & ~C|), the
new element is in the sample labeled $2$ (\verb|~A & B & ~C|),
the new element is in the sample labeled $3$ (\verb|~A & ~B & C|).
Here the tilde is \emph{sympy's} notation for negation, $|$ is disjunction, and
$\&$ is conjunction. The next constraint in line $8$ means that
it is not possible to determine that the sample labeled $2$ 
is empty in pre-testing if it contains the new element.
The following three lines $9-11$ state that without any
additional information the probability of the element being 
each of the three samples is identical. The statement in 
line $12$ means that if the new element is in the sample labeled
$1$ then
it is equally likely that pre-testing will exclude samples 
labeled $2$, $3$. The last statement about probability
in line $13$ means that if the new element is in sample labeled $3$
then the sample labeled $2$ will certainly be determined
to be empty in pre-testing. Finally in line $14$ the 
\emph{p} object is finalized excluding the possibility
of adding more constraints and statements about probability.

Next in lines $16-21$ the probability vector describing the
hypothetical theoretical scenario is calculated. In line $16$ the joint matrix $(A|b)$ is calculated and the linear system $A x = b$
is solved for $x$ in line $21$. Note that the variable name
\emph{x} has no relation to the symbol $x$ we use for implicit assumptions. 
Additionally, we are using the
\emph{np.linalg.lstsq} function because the main matrix of the linear
system is rectangular.

With the solution $x$ we can work out the values of
probabilities $P(u | v x)$ where $u$, $v$ are arbitrary
logical statements (constructed from the literals $A$, $B$, $C$, and $O$) in the system using the property from \eqref{numdem}. 
The function \emph{getNumDem} in lines $23$ and $28$ returns
vectors containing $u_{d} v_{d}$ and $v_{d}$ from the numerator
and denominator in \eqref{numdem}. After summing these vectors
with the probabilities in lines $24$ and $29$ the 
resulting probability values can be printed to standard output.

\subsection{Maximizing entropy}

The \emph{examples} folder in \cite{TopolnickiGitHub} contains 
further demonstrations. For example, additional logical statements
$D$, $E$, $F$ may be added that are irrelevant to the considered
hypothetical scenario in order to test entropy maximization. However, it turns out 
that the linear program described 
in Section \ref{max} returns probability vectors that 
are already very close to the maximum entropy configuration.

To test the entropy maximizing procedure we will deliberately 
spoil the solution and run it through the optimization procedure:
\begin{lstlisting}[language=Python , numbers = left]
from piter import Piter
from sympy.abc import d , m , a , b , c , e , f , g , h
from sympy import true
import numpy as np

p = Piter({d , m , a , b , e , f , g , h})
p.addP(e & ~a & ~b & ~d & ~f & ~m & g & h, true , 0.02)
p.addP(b & f & m & ~a & ~d & ~e & g & h , true , 0.02)
p.finalize()

vv , ns = p.getPositiveSolition()

other = vv > 0.01
newvv = np.random.rand(vv.shape[0])
newvv[other] = 0.0
newvv = 0.96 * newvv / np.sum(newvv)
newvv[other] = vv[other]

print("this schould be 3 : " , np.sum(other))
print("this schould be close to 1 : " , np.sum(newvv[:-1]))

np.savetxt("iterations_0" , newvv[:-1])
res80 = p.optimizeEntropy(vv , ns , epochs = 80)
np.savetxt("iterations_80" , res80)
res160 = p.optimizeEntropy(vv , ns , epochs = 160)
np.savetxt("iterations_160" , res160)
res320 = p.optimizeEntropy(vv , ns , epochs = 320)
np.savetxt("iterations_320" , res320)
\end{lstlisting}

In this code lines $1-4$ contain imports similar to those in
Subsection \ref{ssol} but the number of symbols imported
from \emph{sympy.abc} is greater and will result in 
$256$ dimensional probability vectors. The logic
is simpler, lines $6-9$ state that only two specific
values of probabilities are specified and each has a
value of $0.02$. The positive solution for the probability
vector is calculated in line $11$ using a linear program.
The \emph{vv} vector contains the probability values supplemented
by the value $\alpha = 1$ at the end and
the \emph{ns} contains the null space vectors.

\begin{figure}[h!]
	\centering
	\includegraphics[width = 1.0 \textwidth]{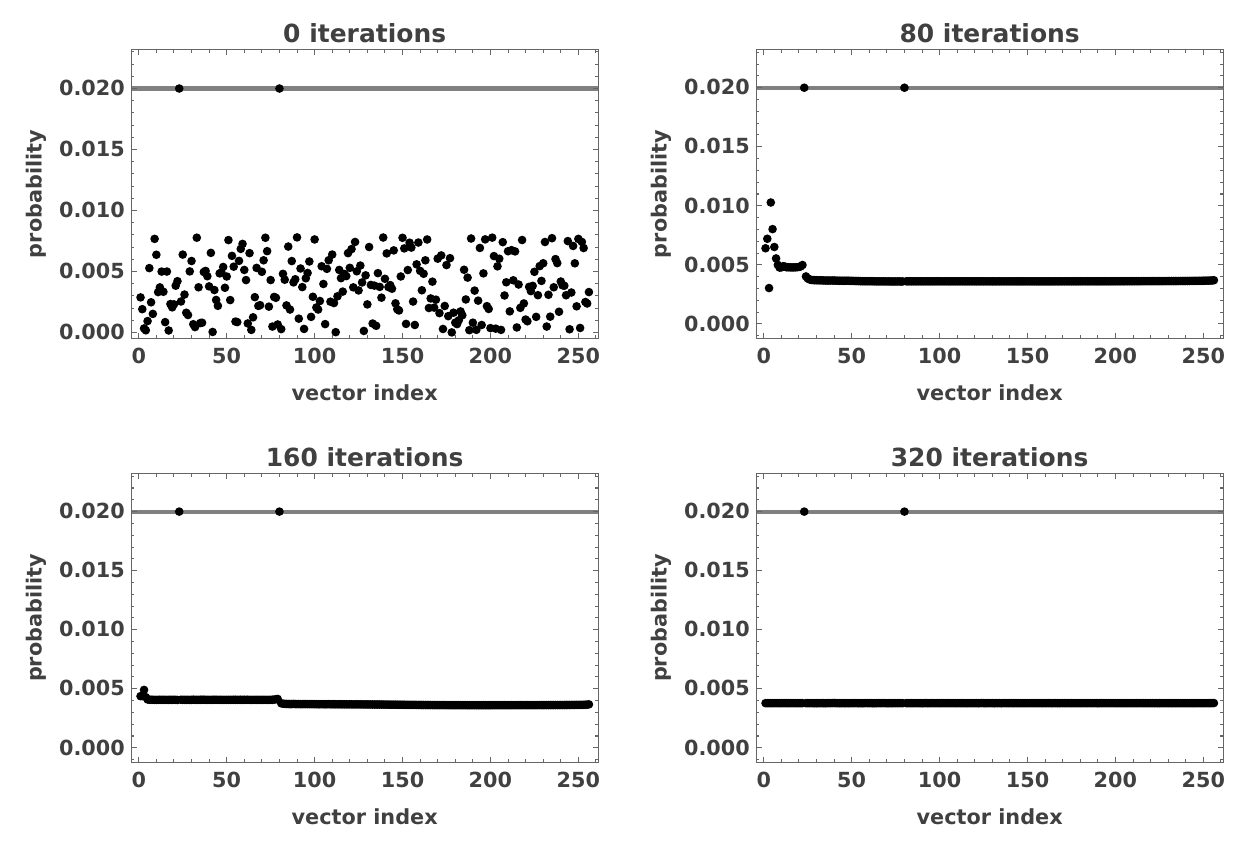}
	\caption{Maximizing entropy in consecutive iterations of the procedure
    from Section \ref{max}. The top left plot shows the initial vector.
    The probability values other then $0.02$ gradually tend tward a 
    plateou with an increasing number of iterations.}
	\label{iters}
\end{figure}

Next in lines $13-17$ the positive solution 
is artificially spoiled. We leave the vector components
that correspond to $0.02$ the same in \emph{vvnew} as
in \emph{vv}. The remainig values are drawn from a uniform
distribution. Instructions in lines $19-20$ can be 
used to verify that this process was succesfull. 
Finally in lines $22-32$ we run the entropy optimization
procedure using the spoiled vector and the same
null space vectors \emph{ns}. Results for
$0$, $80$, $160$, and $320$ iterations (using 
the \emph{Adam} optimizer \cite{kingma2017adammethodstochasticoptimization} and learning rate $0.01$)
are illustrated in Figure \ref{iters}. With an increasing number 
of iterations, the probabilities other than the two $0.02$ values
gradually tend to a plateau. After $320$ iterations, the 
probabilities are essentially flat as would be expected
from an entropy maximizing configuration.

The \emph{examples} folder in \cite{TopolnickiGitHub} contains 
additional sample calculations. This includes versions of the Monty Hall problem that
test whether the computed solution is influenced by adding logical statements that are not relevant to the problem. These additional statements significantly 
increase the size of the problem, but do not change the final result of 
the calculation. 
The \emph{tests} folder in \cite{TopolnickiGitHub} contains unit tests
written in part based on these examples.

\section{Summary and outlook}
\label{so}

This paper presented an approach to 
describe systems governed by discrete probabilities.
These description is specified by very general probability 
statements \eqref{input} and 
the methods described in section \ref{cons}, \ref{max}
together with the implementation available in \cite{TopolnickiGitHub}
and code examples in Section \ref{aions} can be used to 
work out probabilities in these systems. The implementation 
\cite{TopolnickiGitHub} is a direct reflection of the theoretical
framework described in this paper, making it possible to easily
modify and extend the code.

A description of physical systems in terms of discrete probabilities
is suitable for a wide range of phenomena.
While there are existing tools for working with continuous distributions
there is, to the best of our knowledge, a lack of software solutions
designed for discrete probabilities. As demonstrated in 
Subsection \ref{ssol} calculations of this nature are often
unintuitive and obtaining a correct solution, even in simple cases,
may be challenging. For more complicated problem a tool based
on the framework described in this paper may be valuable.

The methods desribed in this paper may also have applications
outside of physics. It would be very interesting to explore 
the application of the new library \cite{TopolnickiGitHub}
to the analysis of journalistic texts in the context of
probability correctness. A first step in this direction
was made in the linter program \cite{SikoraGitHub}.
The code in this repository uses \cite{TopolnickiGitHub}
to implement a linter capable of analyzing natural text
and verify if the text contains incorrect statements related
to probability. We believe that spotting errors of this 
type in journalistic text and on social media is a step 
towards reducing misinformation.

\section*{Acknowledgments}

This work was supported by the National Science Center,
Poland under Grant IMPRESS-U 2024/06/Y/ST2/00135. 

\section*{Data Availability Statement}

The python library and code necessary to reproduce the 
results presented in this paper is available in the
public \emph{GitHub} repository 
\url{https://github.com/kacpertopolnicki/ProbabilityIter}.

\bibliography{bibl}

@book{cox1961algebra,
  title = {Algebra of Probable Inference},
  ISBN = {9780801869822},
  url = {http://dx.doi.org/10.56021/9780801869822},
  DOI = {10.56021/9780801869822},
  publisher = {Johns Hopkins University Press},
  author = {Cox,  Richard},
  year = {2001}
}

@book{jaynes2003probability,
  title = {Probability Theory: The Logic of Science},
  ISBN = {9780511790423},
  url = {http://dx.doi.org/10.1017/CBO9780511790423},
  DOI = {10.1017/cbo9780511790423},
  publisher = {Cambridge University Press},
  author = {Jaynes,  E. T.},
  editor = {Bretthorst,  G. Larry},
  year = {2003},
  month = Apr 
}

@book{kolmogorov1950foundations,
  title     = "Foundations of the theory of probability",
  author    = "Kolmogorov, A N",
  publisher = "Dover Publications",
  month     =  mar,
  year      =  2018,
  address   = "New York, NY",
  language  = "en"
}

@article{10.7717/peerj-cs.103,
     title = {SymPy: symbolic computing in Python},
     author = {Meurer, Aaron and Smith, Christopher P. and Paprocki, Mateusz and \v{C}ert\'{i}k, Ond\v{r}ej and Kirpichev, Sergey B. and Rocklin, Matthew and Kumar, AMiT and Ivanov, Sergiu and Moore, Jason K. and Singh, Sartaj and Rathnayake, Thilina and Vig, Sean and Granger, Brian E. and Muller, Richard P. and Bonazzi, Francesco and Gupta, Harsh and Vats, Shivam and Johansson, Fredrik and Pedregosa, Fabian and Curry, Matthew J. and Terrel, Andy R. and Rou\v{c}ka, \v{S}t\v{e}p\'{a}n and Saboo, Ashutosh and Fernando, Isuru and Kulal, Sumith and Cimrman, Robert and Scopatz, Anthony},
     year = 2017,
     month = jan,
     volume = 3,
     pages = {e103},
     journal = {PeerJ Computer Science},
     issn = {2376-5992},
     url = {https://doi.org/10.7717/peerj-cs.103},
     doi = {10.7717/peerj-cs.103}
    }

@MISC {5128410,
    TITLE = {Transformation to basis with all positive vectors},
    AUTHOR = {Ben Grossmann (https://math.stackexchange.com/users/81360/ben-grossmann)},
    HOWPUBLISHED = {Mathematics Stack Exchange},
    NOTE = {URL:https://math.stackexchange.com/q/5128410 (version: 2026-03-20)},
    EPRINT = {https://math.stackexchange.com/q/5128410},
    URL = {https://math.stackexchange.com/q/5128410}
}

@inproceedings{DBLP:journals/corr/KingmaB14,
  author       = {Diederik P. Kingma and
                  Jimmy Ba},
  editor       = {Yoshua Bengio and
                  Yann LeCun},
  title        = {Adam: {A} Method for Stochastic Optimization},
  booktitle    = {3rd International Conference on Learning Representations, {ICLR} 2015,
                  San Diego, CA, USA, May 7-9, 2015, Conference Track Proceedings},
  year         = {2015},
  url          = {http://arxiv.org/abs/1412.6980},
  bibsource    = {dblp computer science bibliography, https://dblp.org}
}

@ARTICLE{2020SciPy-NMeth,
  author  = {Virtanen, Pauli and Gommers, Ralf and Oliphant, Travis E. and
            Haberland, Matt and Reddy, Tyler and Cournapeau, David and
            Burovski, Evgeni and Peterson, Pearu and Weckesser, Warren and
            Bright, Jonathan and {van der Walt}, St{\'e}fan J. and
            Brett, Matthew and Wilson, Joshua and Millman, K. Jarrod and
            Mayorov, Nikolay and Nelson, Andrew R. J. and Jones, Eric and
            Kern, Robert and Larson, Eric and Carey, C J and
            Polat, {\.I}lhan and Feng, Yu and Moore, Eric W. and
            {VanderPlas}, Jake and Laxalde, Denis and Perktold, Josef and
            Cimrman, Robert and Henriksen, Ian and Quintero, E. A. and
            Harris, Charles R. and Archibald, Anne M. and
            Ribeiro, Ant{\^o}nio H. and Pedregosa, Fabian and
            {van Mulbregt}, Paul and {SciPy 1.0 Contributors}},
  title   = {{{SciPy} 1.0: Fundamental Algorithms for Scientific
            Computing in Python}},
  journal = {Nature Methods},
  year    = {2020},
  volume  = {17},
  pages   = {261--272},
  adsurl  = {https://rdcu.be/b08Wh},
  doi     = {10.1038/s41592-019-0686-2},
}

@article{pymc2023,
  title = {{PyMC}: A Modern and Comprehensive Probabilistic Programming Framework in {P}ython},
  author = {Oriol Abril-Pla and Virgile Andreani and Colin Carroll and Larry Dong and Christopher J. Fonnesbeck and Maxim Kochurov and Ravin Kumar and Junpeng Lao and Christian C. Luhmann and Osvaldo A. Martin and Michael Osthege and Ricardo Vieira and Thomas Wiecki and Robert Zinkov },
  journal = {{PeerJ} Computer Science},
  volume = {9},
  number = {e1516},
  doi = {10.7717/peerj-cs.1516},
  year = {2023}
}

@article{pyro_joss,
         doi = {10.21105/joss.01265},
         url = {https://doi.org/10.21105/joss.01265},
         year = {2019},
         publisher = {The Open Journal},
         volume = {4},
         number = {34},
         pages = {1265},
         author = {Alice Harpole and Michael Zingale and Ian Hawke and Taher Chegini},
         title = {pyro: a framework for hydrodynamics explorations and prototyping},
         journal = {Journal of Open Source Software}
}

@misc{TopolnickiGitHub,
  author = {Kacper Topolnicki},
  title = {ProbabilityIter},
  year = {2026},
  publisher = {GitHub},
  journal = {GitHub repository},
  howpublished = {\url{https://github.com/kacpertopolnicki/ProbabilityIter}},
  url = {https://github.com/kacpertopolnicki/ProbabilityIter}
}

@misc{SikoraGitHub,
  author = {Krystian Sikora},
  title = {ProbabilityLinter},
  year = {2026},
  publisher = {GitHub},
  journal = {GitHub repository},
  howpublished = {\url{https://github.com/krystian-sikora/ProbabilityLinter}},
  url = {https://github.com/krystian-sikora/ProbabilityLinter}
}

@Inbook{Smith1989,
author="Smith, C. Ray
and Erickson, Gary",
editor="Skilling, J.",
title="From Rationality and Consistency to Bayesian Probability",
bookTitle="Maximum Entropy and Bayesian Methods: Cambridge, England, 1988",
year="1989",
publisher="Springer Netherlands",
address="Dordrecht",
pages="29--44",
isbn="978-94-015-7860-8",
doi="10.1007/978-94-015-7860-8_2",
url={https:\/\/doi.org\/10.1007\/978-94-015-7860-8\_2}
}

@misc{kingma2017adammethodstochasticoptimization,
      title={Adam: A Method for Stochastic Optimization}, 
      author={Diederik P. Kingma and Jimmy Ba},
      year={2017},
      eprint={1412.6980},
      archivePrefix={arXiv},
      primaryClass={cs.LG},
      url={https://arxiv.org/abs/1412.6980}, 
}

@book{granberg2014monty,
  title={The Monty Hall Dilemma: A Cognitive Illusion Par Excellence},
  author={Granberg, D.},
  isbn={9780996100809},
  lccn={2014906811},
  url={https://books.google.pl/books?id=S3X1rQEACAAJ},
  year={2014},
  publisher={Lumad Publishing}
}

@misc{mh,
  title = {The Monty Hall Problem},
  author = {Afra Zomorodian},
  howpublished = {\url{https://bechtel.colorado.edu/~balajir/CVEN5454/lectures/monty.pdf}},
  note = {Accessed: 2010-09-30}
}

\end{document}